%% file: main.tex
\documentclass[11pt,letterpaper]{article}

\input{_packages}

\input{_format}

\input{_macros}

\begin{document}

\includepdf[pages=-]{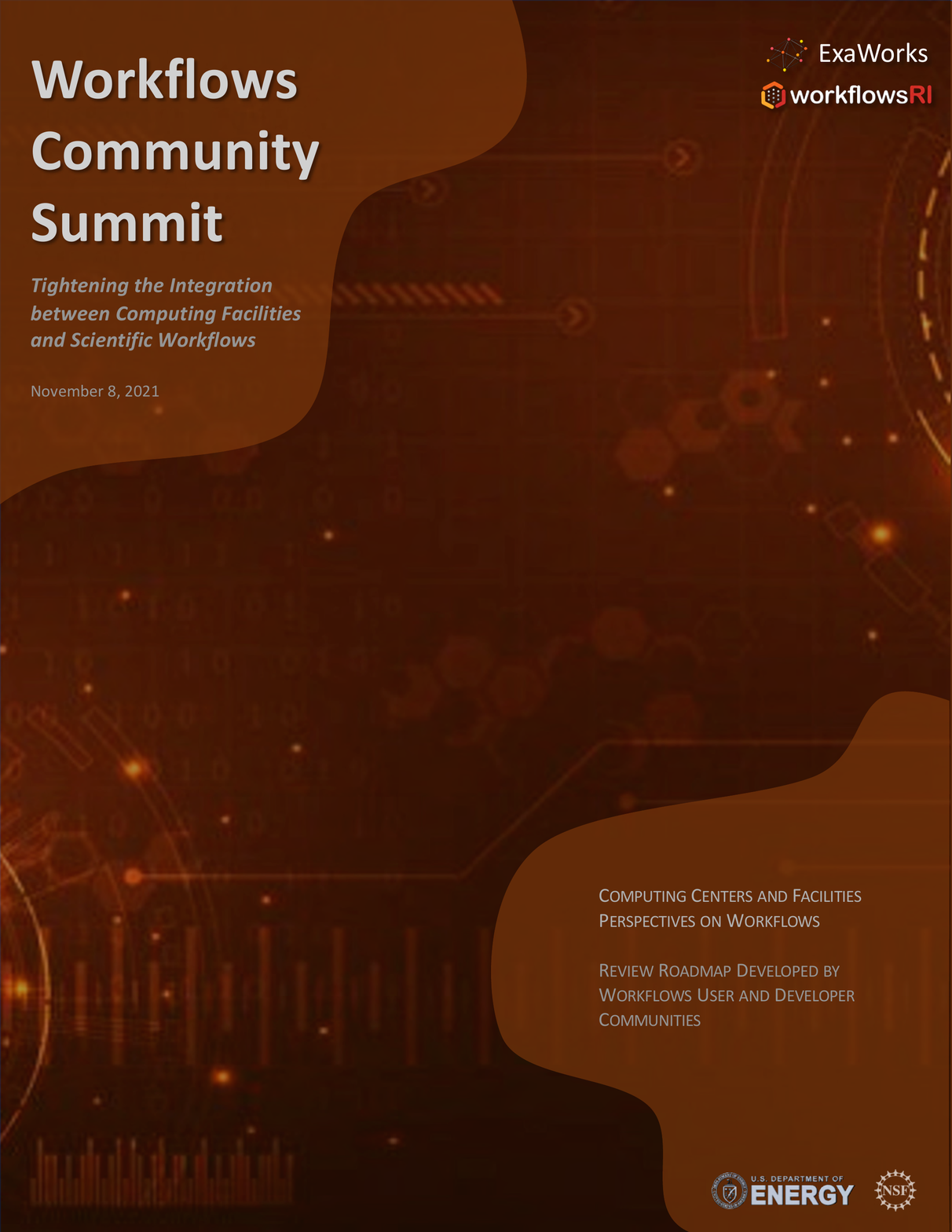}


\input{_headers}

\input{frontmatter}
\newpage


\newpage


\input{introduction}

\newpage
\section{Summit Discussions}

\input{review}

\input{perspectives}


\section{Looking Forward}
\input{summary}

\input{_references}

\newpage
\input{appx_contributors}
\newpage
\input{appx_agenda}

\end{document}

%% file: _packages.tex
\usepackage{times}
\usepackage{relsize}
\usepackage{microtype}
\usepackage{cite}
\usepackage{xspace}
\usepackage[table,xcdraw]{xcolor}
\usepackage{color,colortbl}
\usepackage{subfigure}
\usepackage{alltt}
\usepackage{url}
\usepackage{comment}
\usepackage{graphicx}
\usepackage{relsize}
\usepackage{paralist}
\usepackage{todonotes}
\usepackage{wrapfig}
\usepackage{titlesec}
\usepackage{epsfig}
\usepackage{geometry}
\usepackage{tikz}
\usepackage[normalem]{ulem}
\usepackage{titlesec}
\usepackage{booktabs}
\usepackage{wrapfig}
\usepackage{multirow}
\usepackage{pifont}
\usepackage{wrapfig}
\usepackage{soul}
\usepackage{enumitem}
\usepackage{tcolorbox}
\usepackage{pdfpages}
\usepackage{fancyhdr}
\usepackage{longtable}
\usepackage{makecell}
\usepackage{lettrine}
\usepackage{listings}
\usepackage{longtable}

%% file: _format.tex
\titlespacing{\section}{0pc}{1pc}{1pc}
\titlespacing{\subsection}{0pc}{2pc}{1pc}
\titlespacing{\subsubsection}{0pc}{1pc}{1pc}
\titleformat*{\section}{\Large\bfseries}
\titleformat*{\subsection}{\large\bfseries}
\titleformat*{\subsubsection}{\normalsize\bfseries}

\lstdefinestyle{wcsStyle}{
  tabsize=4,
  showspaces=false,
  showstringspaces=false,
  aboveskip=0em,
  belowskip=0em,
}

%% file: _macros.tex
\definecolor{Gray}{gray}{0.9}

\newcommand{\pp}[1]{\medskip \noindent \textbf{#1.}\xspace}

\newcommand{\bp}[2]{\begin{tabular}{@{\textbullet~}p{#1}@{}}#2\end{tabular}}

%% file: _headers.tex
\pagestyle{fancy}
\fancyhf{}
\rhead{
  \includegraphics[height=11pt]{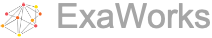} \ \ 
  \includegraphics[height=12pt]{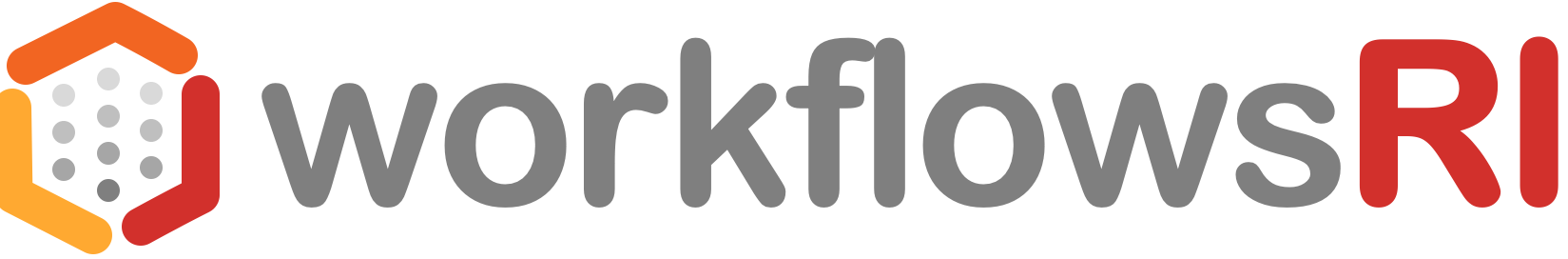} 
}
\lhead{Workflows Community Summit - November 2021}
\rfoot{\thepage}

%% file: frontmatter.tex
\begin{table}[!ht]
\centering
\smaller
\begin{tabular}{p{16cm}}
    \vspace{2em}
    \textbf{Disclaimer}
    \\
    The Workflows Community Summit was supported by the National Science Foundation (NSF) under grants number 2016610, 2016619, and 2016682, and by the Exascale Computing Project (17-SC-20-SC), a collaborative effort of the U.S. Department of Energy Office of Science and the National Nuclear Security Administration. This research used resources of the Oak Ridge Leadership Computing Facility at the Oak Ridge National Laboratory, which is supported by the Office of Science of the U.S. Department of Energy under Contract No. DE-AC05-00OR22725.
    \\ \medskip
    This report was prepared as an account of work sponsored by agencies of the United States Government. Neither the United States Government nor any agency thereof, nor any of their employees, makes any warranty, express or implied, or assumes any legal liability or responsibility for the accuracy, completeness, or usefulness of any information, apparatus, product, or process disclosed, or represents that its use would not infringe privately owned rights. Reference herein to any specific commercial product, process, or service by trade name, trademark, manufacturer, or otherwise, does not necessarily constitute or imply its endorsement, recommendation, or favoring by the United States Government or any agency thereof. The views and opinions of authors expressed herein do not necessarily state or reflect those of the United States Government or any agency thereof.
    \\
    \vspace{0.5em}
    \textbf{Preferred citation} 
    \\
    R. Ferreira da Silva, K. Chard, H. Casanova, D. Laney, D. Ahn, S. Jha, W. E. Allcock, G. Bauer, D. Duplyakin, B. Enders, T. M. Heer, E. Lançon, S. Sanielevici, K. Sayers,
    ``Workflows Community Summit: Tightening the Integration between Computing Facilities and Scientific Workflows", Technical Report, January 2022, DOI: 10.5281/zenodo.5815332.
    \\
    \rowcolor[HTML]{F7F7F7}
    \lstset{basicstyle=\scriptsize,style=wcsStyle}
    \begin{lstlisting}
@misc{wcs2021facilities,
  author    = {Ferreira da Silva, Rafael and Chard, Kyle and Casanova, Henri and Laney, Dan 
               and Ahn, Dong and Jha, Shantenu and Allcock, William E. and Bauer, Gregory
               and Duplyakin, Dmitry and Enders, Bjoern and Heer, Todd M. and Lan\c{c}on, 
               Eric and Sanielevici, Sergiu and Sayers, Kevin},
  title     = {{Workflows Community Summit: Tightening the Integration between Computing 
               Facilities and Scientific Workflows}},
  month     = {January},
  year      = {2022},
  publisher = {Zenodo},
  doi       = {10.5281/zenodo.5815332}
}
    \end{lstlisting}
    \\
    \vspace{0.5em}
    \textbf{License}
    \\
    This report is made available under a Creative Commons Attribution-ShareAlike 4.0 International license ({\scriptsize \url{https://creativecommons.org/licenses/by-sa/4.0/}}).
\end{tabular}
\end{table}
\vspace*{\fill}

%% file: introduction.tex
\section{Introduction}
\label{sec:introduction}

Scientific workflows are used almost universally across science domains for solving complex and large-scale computing and data analysis problems. The importance of workflows is highlighted by the fact that they have underpinned some of the most significant discoveries of the past decades~\cite{badia2017workflows}. Many of these workflows have significant computational, storage, and communication demands, and thus must execute on a range of large-scale computer systems, from local clusters to public clouds and upcoming exascale HPC platforms~\cite{ferreiradasilva-fgcs-2017}. Managing these executions is often a significant undertaking, requiring a sophisticated and versatile software infrastructure.

Historically, infrastructures for workflow execution consisted of complex, integrated systems, developed in-house by workflow practitioners with strong dependencies on a range of legacy technologies---even including sets of ad hoc scripts. Due to the increasing need to support workflows, dedicated workflow systems were developed to provide abstractions for creating, executing, and adapting workflows conveniently and efficiently while ensuring portability.  While these efforts are all worthwhile individually, there are now hundreds of independent workflow systems~\cite{workflow-systems}. These workflow systems are created and used by thousands of researchers and developers, leading to a rapidly growing corpus of workflows research publications. The resulting workflow system technology landscape is fragmented, which may present significant barriers for future workflow users due to many seemingly comparable, yet usually mutually incompatible, systems that exist.

In order to tackle some of the challenges described above, the DOE-funded ExaWorks~\cite{exaworks} and NSF-funded WorkflowsRI~\cite{workflowsri} projects have organized in 2021 a series of events entitled the ``Workflows Community Summit". The third edition of the ``Workflows Community Summit" explored workflows challenges and opportunities from the perspective of computing centers and facilities. This third summit builds on two prior summits (\url{https://workflowsri.org/summits}) that (i)~established a high level vision for workflows research; and (ii)~explored technical approaches for realizing that vision~\cite{ferreiradasilva2021wcs, wcs2021technical}. The third summit brought together a small group of facilities representatives with the aim to understand how workflows are currently being used at each facility, how facilities would like to interact with workflow developers and users, how workflows fit with facility roadmaps, and what opportunities there are for tighter integration between facilities and workflows.

This report documents and organizes the wealth of information provided by the participants before, during, and after the summit.

\subsection{Summit Organization}

\begin{figure}[!t]
    \centering
    \includegraphics[width=\linewidth]{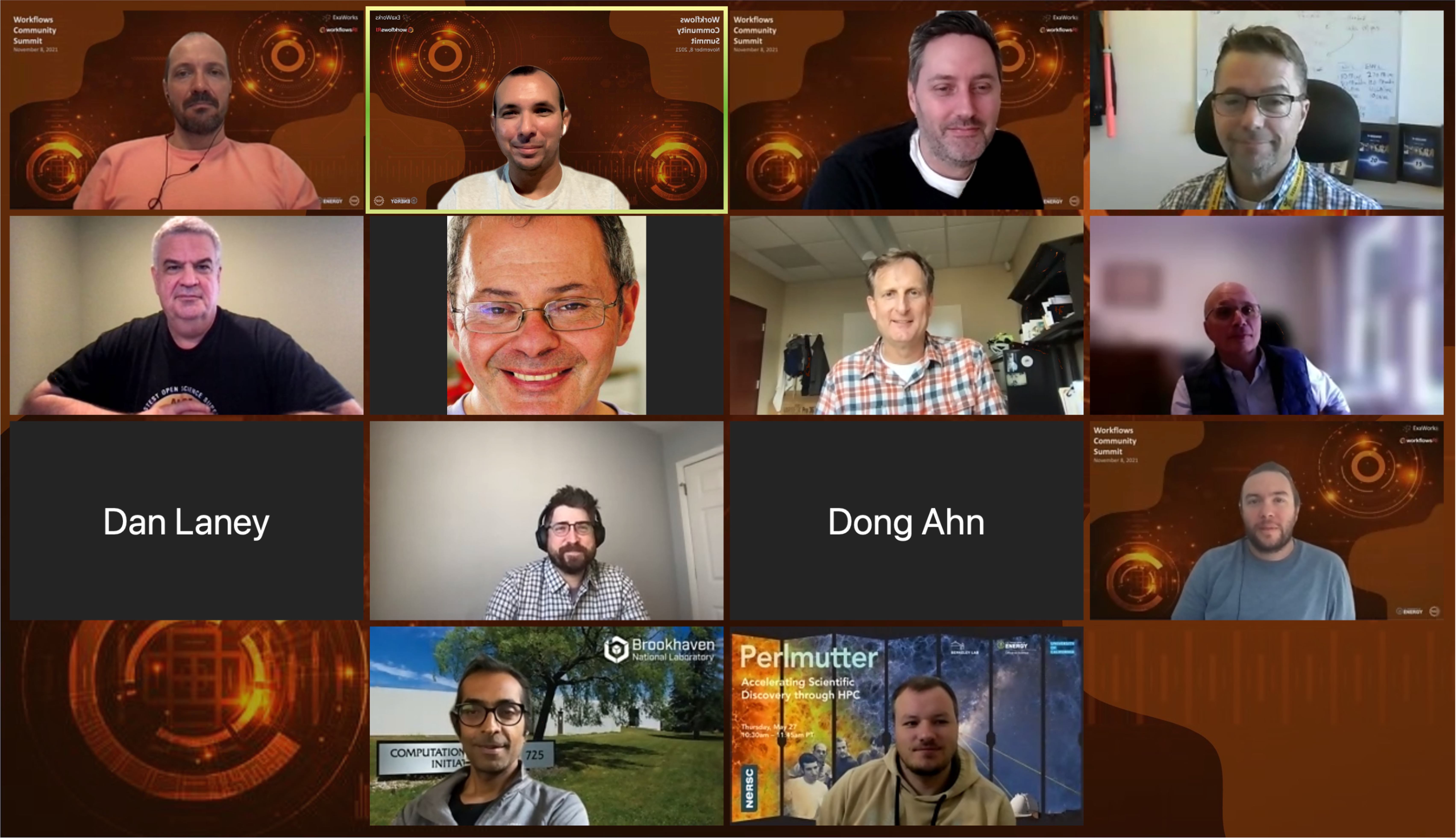}
    \caption{Screenshot of the third edition of the Workflows Community Summit participants. (The event was held virtually via Zoom on November 8, 2021.)}
    \label{fig:participants}
\end{figure}

This document reports on discussions and findings from the third edition of a ``Workflows Community Summit" that took place on November 8, 2021~\cite{wcs}.
The summit included 14 participants from various computing centers and facilities (Figure~\ref{fig:participants}):

\begin{compactitem}
    \item ALCF: Argonne Leadership Computing Facility
    \item BNL: Brookhaven National Laboratory
    \item LLNL: Lawrence Livermore National Laboratory
    \item NCSA: National Center for Supercomputing Applications
    \item NERSC: National Energy Research Scientific Computing Center
    \item NREL: National Renewable Energy Laboratory
    \item OLCF: Oak Ridge Leadership Computing Facility
    \item PSC: Pittsburgh Supercomputing Center
\end{compactitem}

\begin{table}[!t]
    \centering
    \small
    \setlength{\tabcolsep}{12pt}
    \begin{tabular}{lll}
        \toprule
        \textbf{Project} & \textbf{Name} & \textbf{Affiliation} \\
        \midrule
        WorkflowsRI & Rafael Ferreira da Silva & Oak Ridge National Laboratory \\
        & Kyle Chard & University of Chicago \\
        & Henri Casanova & University of Hawai'i at M\~anoa \\
        \rowcolor[HTML]{F2F2F2}
        ExaWorks & Dan Laney & Lawrence Livermore National Laboratory \\
        \rowcolor[HTML]{F2F2F2}
        & Dong H. Ahn & Lawrence Livermore National Laboratory \\
        \rowcolor[HTML]{F2F2F2}
        & Kyle Chard & Argonne National Laboratory\\
        \rowcolor[HTML]{F2F2F2}
        & Shantenu Jha & Brookhaven National Laboratory \\
        \bottomrule
    \end{tabular}
    \caption{Workflows Community Summit organizers.}
    \label{tab:organization}
\end{table}

\medskip

The summit was co-organized by the PIs of the ExaWorks and WorkflowsRI projects. These two projects aim to develop solutions to address some of the challenges faced by the workflow community, albeit in different ways.

\pp{ExaWorks}
Aims to develop a multi-level workflows SDK that will enable teams to produce scalable and portable workflows for a wide range of exascale applications. ExaWorks does not aim to replace the many workflow solutions already deployed and used by scientists, but rather to provide a robust SDK and work with the community to identify well-defined and scalable components and interfaces that can be leveraged by new and existing workflows. A key goal is that SDK components should be designed to be usable by many other workflow management systems, thus facilitating software convergence in the workflows community.

\pp{WorkflowsRI}
Focuses on defining the blueprint for a community research and development infrastructure. Specifically, the goal is to bring together workflow users, developers, and researchers to identify common ways to develop, use, and evaluate workflow systems, resulting in a useful, maintainable, and community-vetted set of principles, methods, and metrics for workflow systems research and development. The resulting research infrastructure will provide an academic nexus for addressing pressing workflow challenges, including those resulting from the increasing heterogeneity and size of computing systems, changing workflow patterns to encompass rapid feedback and machine learning models, and the increasing diversity of the user community.

\subsection{Summit Structure and Activities}

Based on the outcomes of the first two summits, we identified the need for sharing these outcomes with computing centers and facilities, as well as capturing their perspectives regarding workflow systems and applications. To this end, this third summit was organized as a focused session with a small group of participants as follows:

\begin{compactenum}
    \item Summary of previous Workflows Community Summits and their outcomes.
    \item Discussion of facility perspectives on workflows including questions such as:
    \begin{compactitem}
        \item What is a workflow?
        \item What are some of the challenges you see supporting workflows?
        \item What are your current practices to support workflows (e.g., documentation, native systems, etc.)?
    \end{compactitem}
    \item Brief introduction to the ExaWorks project
    \item Review of the roadmap developed by workflows user and developer communities
\end{compactenum}

\medskip \noindent
In the following sections, we summarize the outcome of these discussions.

%% file: review.tex
\subsection{A Community Roadmap for Scientific Workflows Research and Development}
\label{sec:review}

\input{tbl_challenges_efforts}

The overarching goal of the two first editions of the ``Workflows Community Summits" was to (i)~develop a view of the state of the art, (ii)~identify key research challenges, (iii)~articulate a vision for potential activities, and (iv)~explore technical approaches for realizing (part of) this vision. The summits gathered over 70 participants, including lead researchers and developers from around the world, and spanning distinct workflow systems and user communities. 
The outcomes of the previous two summits have been compiled and published in two technical reports~\cite{ferreiradasilva2021wcs, wcs2021technical}.
We synthesized the discussions and findings into a community roadmap~\cite{ferreiradasilva2021works} that presents a consolidated view of the state of the art, challenges, and  potential activities.
Table~\ref{tab:challenges} presents, in the form of top-level themes, a summary of those challenges and targeted community activities. Table~\ref{tab:roadmap} summarizes a proposed community roadmap with technical approaches. 

\input{tbl_roadmap}

%% file: tbl_challenges_efforts.tex
\begin{table*}[!b]
\centering
\scriptsize
\begin{tabular}{p{1.3cm}p{6.9cm}p{6.9cm}}
\toprule 
Theme & Challenges & Community Activities \\
\midrule
\makecell[l]{FAIR\\Computational\\Workflows} &
\bp{6.9cm}{
    FAIR principles for computational workflows that consider the complex lifecycle from specification to execution and data products
    \\
    Metrics to measure the ``FAIRness'' of a workflow
    \\
    Principles, policies, and best practices
}
&
\bp{6.9cm}{
    Review prior and current efforts for FAIR data and software with respect to workflows, and outline principles for FAIR workflows
    \\
    Define recommendations for FAIR workflow developers and systems
    \\
    Automate FAIRness in workflows by recording necessary provenance data
}
\\
\midrule

\makecell[l]{AI\\Workflows} &
\bp{6.9cm}{
    Support for heterogeneous compute resources and fine-grained data management features, versioning, and data provenance capabilities
    \\
    Capabilities for enabling workflow steering and dynamic workflows
    \\
    Integration of ML frameworks into the current HPC landscape
}
&
\bp{6.9cm}{
    Develop comprehensive use cases for sample problems with representative workflow structures and data types
    \\
    Define a process for characterizing the challenges for enabling AI workflows
    \\
    Develop AI workflows as a way to benchmark HPC systems
}
\\
\midrule

\makecell[l]{Exascale\\Challenges\\and Beyond} &
\bp{6.9cm}{
    Resource allocation policies and schedulers are not designed for workflow-aware abstractions, thus users tend to use an ill-fitted job abstraction
    \\
    Unfavorable design of resource descriptions and mechanisms for workflow users/systems, and lack of fault-tolerance and fault-recovery solutions
}
&
\bp{6.9cm}{
    Develop documentation in the form of workflow templates/recipes/miniapps for execution on high-end HPC systems
    \\
    Specify benchmark workflows for exascale execution
    \\
    Include workflow requirements as part of the machine procurement process
}
\\
\midrule

\makecell[l]{APIs, Reuse,\\Interoperability,\\and Standards} &
\bp{6.9cm}{
    Workflow systems differ by design, thus interoperability at some layers is likely to be more impactful than others
    \\
    Workflow standards are typically developed by a subset of the community
    \\
    Quantifying the value of common representations of workflows is not trivial
}
&
\bp{6.9cm}{
    Identify differences and commonalities between different systems
    \\
    Identify and characterize domain-specific efforts, identify workflow patterns, and develop case-studies of business process workflows and serverless workflow systems
}
\\
\midrule

\makecell[l]{Training\\and Education} &
\bp{6.9cm}{
    Many workflow systems have high barriers to entry and lack training materials
    \\
    Homegrown workflow solutions and constraints can prevent users from reproducing their functionality on workflow systems developed by others
    \\
    Unawareness of the workflow technological and conceptual landscape
}
&
\bp{6.9cm}{
    Identify basic sample workflow patterns, develop a community workflow knowledge-base, and look at current research on technology adoption
    \\
    Include workflow terminology and concepts in university curricula and software carpentry efforts
}
\\
\midrule

\makecell[l]{Building\\a Workflows\\Community} &
\bp{6.9cm}{
    Diverse definitions of a ``workflows community''
    \\
    Remedy the inability to link developers and users to bridge translational gaps
    \\
    Pathways for participation in a network of researchers, developers, and users
}
& 
\bp{6.9cm}{
    Establish a common knowledge-base for workflow technology
    \\
    Establish a \emph{Workflow Guild}: an organization focused on interaction and relationships, providing self-support between workflow developers and their systems
}
\\
\bottomrule
\end{tabular}
\caption{Summary of current workflows research and development challenges and proposed community activities~\cite{ferreiradasilva2021works}.}
\label{tab:challenges}
\end{table*}

%% file: tbl_roadmap.tex
\begin{table*}[!t]
\centering
\scriptsize
\begin{tabular}{p{3.2cm}p{12.3cm}}
\toprule 
Thrust & Roadmap Milestones \\
\midrule
\makecell[l]{Definition of common\\workflow patterns and\\benchmarks} &
\bp{12.3cm}{
    Define small sets of workflow patterns and benchmark deliverables, and implement them using a selected set of workflow systems
    \\
    Investigate automatic generation of patterns and configurable benchmarks (e.g., to enable weak and strong scaling experiments)
    \\
    Establish or leverage a centralized repository to host and curate patterns and benchmarks
}
\\
\midrule

\makecell[l]{Identifying paths toward\\interoperability of workflow\\systems} &
\bp{12.3cm}{
    Define interoperability for different roles, develop a horizontal interoperability (i.e., making interoperable components), and establish a requirements document per abstraction layer
    \\
    Develop real-world workflow benchmarks, use cases for interoperability, and common APIs that represent workflow library components
    \\
    Establish a workflow systems developer community
}
\\
\midrule

\makecell[l]{Improving workflow systems'\\interface with legacy and\\ emerging HPC software and\\hardware stacks} &
\bp{12.3cm}{
    Document a machine-readable description of key properties of widely used sites, and remote authentication needs from the workflow perspective
    \\
    Identify new workflow patterns (e.g., motivated by AI workflows), attain portability across heterogeneous hardware, and develop a registry of execution environment information
    \\
    Organize a community event involving workflow system developers, end users, authentication technology providers, and facility operators
}
\\
\bottomrule
\end{tabular}
\caption{Summary of technical roadmap milestones per research and development thrust~\cite{ferreiradasilva2021works}.}
\label{tab:roadmap}
\end{table*}

%% file: perspectives.tex
\subsection{Facility Perspectives on Workflows}
\label{sec:perspectives}

In this third Workflows Community Summit, representatives from  eight computing centers and facilities (ALCF, BNL, LLNL, NCSA, NERSC, NREL, OLCF, and PSC) had the opportunity to analyze and discuss the challenges, potential efforts, and proposed roadmap (Tables~\ref{tab:challenges} and~\ref{tab:roadmap}) developed by users, developers, and researchers from the workflows community. Discussions were guided through sets of questions that encompassed various topics, such as the fundamental definition of a workflow, how workflows are currently supported by computing centers and facilities, software testing and deployment, and user support and training activities.

\subsubsection{What is a workflow?}

The term \emph{scientific workflow} (or \emph{workflow} for short) is  overloaded. Workflows are typically associated with automation in computing, which may involve orchestrating computational tasks across heterogeneous resources (e.g., HPC, cloud, edge); managing data movement operations over different types of computer networks, protocols, and connectivity; interacting with a wide range of distributed services; among others. Workflows can also be defined as sets of computational tasks (traditionally described as a directed acyclic graph) with data dependencies, or with higher level abstractions describing sets of interactions between services and/or entities (e.g., computing centers, science facilities, etc.). As a result, computational workflows can be simply defined by Linux bash scripts, or complex systems that execute experiment campaigns on distributed systems -- this wide range definition is typically what facilities consider to be workflows.

\subsubsection{Support for workflow systems}
\label{sec:support}

The current landscape for workflow systems at computing centers and facilities is fragmented and provides ad hoc solutions for enabling users to run their workflows on HPC systems. Typically, computing centers and facilities attempt to deploy/support a small set of workflow tools driven by user needs. When a user-preferred tool is not natively available, the user usually installs/configures the specific system in user space, for exapmle, on the login nodes assuming no need for ``superuser" access. Although these approaches have been used extensively over the past two decades, they suffer from several shortcomings: (i)~it is unclear how users should obtain proper support (facility \emph{vs.} workflow developers); (ii)~misuse of workflow systems capabilities may harm the computing environment (e.g., exceeding limits and policies, multiple instances, etc.); and (iii)~there is a lack of visibility (or transparency) into the execution (e.g., what is being run, how to analyze the execution, etc.).

More recently, NERSC has developed a process for evaluating and documenting workflow systems. This process is driven by a specialized working group that attempts to evaluate workflow systems in terms of software requirements and ability to run the systems on site (typically performed by an intern). In this process, members of the working group attempt to engage with developers of workflow systems so that they can provide guidance on the installation/configuration steps. If successful, the next step is the development of a comprehensive, yet practical, documentation on how to use the workflow system within the NERSC environment (this process is also performed in collaboration with the workflow system developers)~\cite{nersc-workflows}. Finally, periodic reviews (based on user-issued tickets) are performed to update the documentation of or decommission a specific workflow system. Inspired by this experience, OLCF has also started a similar process for evaluating, deploying, and documenting workflow systems that are to be natively supported by the facility~\cite{olcf-workflows}. A key addition to the process is to also tie a point of contact (often a workflow system's developer) that is committed to promptly respond to and address issues encountered and reported by users.

Despite the undeniable success of the above process, not every computing center or facility can dedicate personnel to establishing and maintaining this process in the long term. As a result, there is an opportunity for promoting workflows as first-class citizens within these computing centers and facilities and therefore allocating appropriate funding/resources for such efforts; and/or leveraging similar processes into a template/model that could be widely adopted by computing centers and facilities (potentially sharing outcomes of specific engagements among similar facilities). A task force could be established to define this template (preferably with recognition and support from funding agencies).

\subsubsection{Awareness of facility's policies and capabilities}

The current process for developing workflow systems is mostly driven by needs from a specific domain or application. While recently developed workflow tools immediately benefit from the latest advances from software engineering, libraries, and computing systems, long-standing workflow tools tend to adapt their code base to support new requirements, while still providing backwards compatibility to legacy applications and systems. Although the former seamlessly integrates with the latest technologies, it may not directly integrate with current HPC systems -- as system libraries are not updated in a timely manner. Conversely, the latter typically becomes more complex as additional methods and mechanisms are necessary for providing broader support. 

\pp{Software deployment}
Many workflow systems have a large \emph{tree} of software dependencies. Container technologies such as Singularity and Shifter (and more recently Nvidia's NGC) became popular in HPC systems due to their ability to encapsulate most software dependencies and also to foster reproducibility. However, updates and upgrades of system software can take significant time and effort. Alternative solutions such as Spack~\cite{gamblin2015spack}, a package manager for HPC systems, are gaining traction and are now widely supported by facilities. In addition to facilitating the deployment process of workflow tools, developers should also consider specifics of the system architecture. For instance, NCSA's Delta system will support GPU partitioning for smaller workloads, and supporting this level of granularity can significantly improve the performance of workflow executions.

\pp{Access to resources}
In addition to the above challenges, each computing center or facility also define their own sets of policies and software stacks. Furthermore, the current security model based on two-factor authentication impedes remote access. To address these challenges, funding agencies have emphasized the need for the development of a computing and data ecosystem~\cite{nsf2020, doe2021} with standardized APIs for authorization and authentication, data access and management, and computing. Examples of such approaches include NERSC's Superfacility API~\cite{enders2020cross}, which  aims to enable the use of all NERSC resources through automated means using popular development tools and techniques; and the Distributed Computing Data Ecosystem (DCDE) effort~\cite{Shankar2019}, which envisions to provide federated access to a variety of distributed resources within the DOE environment. Although these solutions seem promising, whatever the mechanisms in use are (e.g., OAuth, SAML, or some next generation quantum security system) they need to provide sufficient information so that the facility can perform introspection and apply their own policies -- note that a facility may not trust another authorization mechanism that is external to the facility. (Note that DCDE only targets authentication, thus authorization is still deferred to each facility.)

\pp{Execution model}
Running workflow applications on HPC systems implies working with constraints imposed by the batch scheduler. Some of these constraints may severely impact the efficiency of the workflow execution, e.g., batch schedulers are typically configured to impose a cap on the number of concurrent jobs by a single user tat can be in the system. Therefore, workflow systems need to provision resources using an additional job management layer and manage workflow executions within these resource provisions.  Flux~\cite{ahn2020flux} appears as a promising solution that provides a resource and job management framework that expands the scheduler's view beyond the single dimension of \emph{nodes}. A complementary solution would be to extend current HPC job schedulers with specific features for workflow tasks (e.g., for defining or bypassing specific policies) -- NERSC current provides specialized support for Jupyter-based jobs via the definition of \emph{flags}, and a similar approach could be extended to workflows. Recently, some facilities are experimenting with a novel model for deploying and managing services, in an attempt to lower the deployment and maintenance barrier for scientific software frameworks. In this model, services are deployed externally to the HPC system on a container-based cloud environment (e.g., Slate at OLCF and Spin at NERSC), which then has the ability to remotely launch/manage jobs in the HPC environment.

\subsubsection{Workload characteristics}

Understanding the characteristics of application workloads is paramount for designing the next-generation of HPC systems. The typical approach for characterizing these workloads is to collect telemetry data of current system usage and via the execution of benchmarks. Although this process is currently performed very early on in the machine procurement process for sets of well-known HPC applications, workflows are not considered. (The proposal for NCSA's Delta system included close collaboration with the Science Gateway Community Institute with the main of identifying the requirements of that community -- a similar approach could be adopted for workflows.) There is an urgent need to integrate workflow requirements as early as possible into this process, so that the barriers for enabling workflow execution on the upcoming machines will be lower. However, this effort is  preconditioned on the availability of workflow benchmark specifications, as well as API/scheduler specifications, as outlined above. NERSC-10 is currently investigating the definition of workflow benchmarks, and frameworks such as WfCommons~\cite{coleman2021fgcs} could be leveraged for the definition and development of representative such benchmarks.

\subsubsection{Training and education}

There is a strong need for more, better, and new training and education opportunities for workflow users. Many users ``re-invent the wheel" without reusing software infrastructures and workflow systems that would make their workflow development and execution more convenient, more efficient, easier to evolve, and more portable. This is partly due to the lack of comprehensive and streamlined training materials that would guide users through the process of designing a workflow (besides the typical ``toy" examples provided in tutorials)~\cite{ferreiradasilva2021works}. On the other hand, some users already have their workflow application locked into a specific workflow system that may not be supported on or compatible with a particular HPC system of interest. In such a situation, the user will incur additional burden to port their application to another supported workflow tool or, if possible, utilize another computing facility.

One approach to tackle this challenge is to provide comprehensive, yet practical, documentation (as discussed in Section~\ref{sec:support}), however considerable and sustainable funding are needed. Another approach is to conduct training for different communities, for example NCSA hosted a series of webinars on workflows~\cite{bw-workflows}. An immediate solution would be to leverage current collaborations with workflow systems developers to tailor their tutorials/examples to specific HPC systems. In some situations, users may directly engage with workflow tools developers when facing an issue; however the facility is often unaware of the problem, thus the solution provided by the developer (that could also benefit other users) is never documented. It would thus be helpful to have key personnel from these tools engage with the facilities so that user support/experience could be enhanced.

\subsubsection{Community resources}

As scientific workflows have become mainstream in modern computational science, the workflows community has significantly grown in the past two decades. However, most of the research and development efforts are ad hoc, and few small, focused collaborative efforts have been working in silos to provide specialized solutions. Therefore, the community lacks a common place to gather and showcase the outcomes of these efforts, and more importantly to condense research and development challenges into a common vision from the community. To address this challenge, the recent established Workflows Community Initiative (WCI)~\cite{wci} seeks to bring the workflows community together (users, developers, researchers, and facilities) to provide community resources and capabilities to enable researchers and workflow systems developers to discover software products, related efforts (e.g., task forces, working groups), events, technical reports, etc. and engage in community-wide efforts to tackle workflows grand challenges. WCI is arising as a community-centered effort for gathering and promoting long-standing and recent community-focused efforts. In addition to these activities, institutional communities of practices related to workflows development and research would also help to achieve the vision for having workflows as first class citizens within computing centers and facilities.

%% file: summary.tex
The needs of scientific applications and the rapid evolution of the computing
continuum paradigm will necessitate novel scientific workflows and increase
the importance  and pervasiveness of existing ones. Given the critical
reliance of scientific discovery process on computing facilities, they must
support the implementation and execution of the entire spectrum of workflows.
In fact, existing and future facilities will present increasingly powerful
capabilities and services, in turn engendering increasingly sophisticated
workflows.

In contrast to traditional approaches where the design and software landscape
of facilities and scientific workflows were disjoint design processes, and
post-facto integrated, there is universal agreement that the future facilities
must be designed \textit{ab initio} to support increasingly sophisticated
end-to-end workflows.

This joint WorkflowsRI-ExaWorks summit represents the start of a new
cooperative  and collaborative approach between the facilities and scientific
workflows community. It provided an important forum to supporting an initial
dialogue between major representatives of facilities and leaders of the
scientific workflows community. 

As outlined in the previous section, we have identified specific areas of
missing capabilities, needs, pain points and collaborative opportunities.
These span the range of community building and strengthening activities, to
the development of diverse and specific workflow benchmarks that will enable
the creation of cross-facility inter-operation. This ExaWorks and Workflows-RI
engagement with facilities will grow in both scope, community participation
and ambition. We invite the community to participate in this open process, and welcome ideas and suggestions for growing this activity.

%% file: _references.tex
\newpage
\cleardoublepage\phantomsection\addcontentsline{toc}{section}{References}


%% file: appx_contributors.tex
\cleardoublepage\phantomsection\addcontentsline{toc}{section}{Appendix A: Participants and Contributors}
\section*{Appendix A: Participants and Contributors}
\label{appx:contributors}

\begin{longtable}[!h]{llp{10.5cm}}
\toprule
\textbf{First Name} & \textbf{Last Name} & \textbf{Affiliation} \\
\midrule
Dong        & Ahn               & Lawrence Livermore National Laboratory \\
\rowcolor[HTML]{F2F2F2} 
William E.  & Allcock           & Argonne National Laboratory \\
Gregory     & Bauer             & National Center for Supercomputing Applications \\
\rowcolor[HTML]{F2F2F2} 
Henri       & Casanova          & University of Hawaii at Manoa   \\
Kyle        & Chard             & University of Chicago           \\
\rowcolor[HTML]{F2F2F2} 
Dimitry     & Duplyakin         & National Renewable Energy Laboratory \\
Bjoern      & Enders            & National Energy Research Scientific Computing Center \\
\rowcolor[HTML]{F2F2F2} 
Rafael      & Ferreira da Silva & Oak Ridge National Laboratory \\
Todd M.     & Herr              & Lawrence Livermore National Laboratory \\
\rowcolor[HTML]{F2F2F2} 
Shantenu    & Jha               & Brookhaven National Laboratory / Rutgers \\
Eric        & Lan\c{c}on        & Brookhaven National Laboratory \\
\rowcolor[HTML]{F2F2F2} 
Daniel      & Laney             & Lawrence Livermore National Laboratory \\
Sergiu      & Sanielevici       & Pittsburgh Supercomputing Center \\
\rowcolor[HTML]{F2F2F2} 
Kevin       & Sayers            & National Renewable Energy Laboratory \\

\bottomrule                                       
\end{longtable}

%% file: appx_agenda.tex
\cleardoublepage\phantomsection\addcontentsline{toc}{section}{Appendix B: Agenda}
\section*{Appendix B: Agenda}
\label{appx:agenda}

\begin{table}[!h]
    \begin{tabular}{lp{12cm}}
        \toprule
        \textbf{Time} & \textbf{Topic} \\
        \midrule
        10:30-10:45am PST & \makecell[l]{\textbf{Welcome and Introductions}\\
        \emph{\small Rafael Ferreira da Silva (Oak Ridge National Laboratory)}} \\
        
        \rowcolor[HTML]{EEEEEE}
        10:45-11:45am PST & \textbf{Facility Perspectives on Workflows}
            \begin{compactitem}
                \item What is a workflow for you?
                \item What are some of the challenges you see supporting workflows?
                \item What are your current practices to support workflows (e.g., documentation, native systems, etc.)?
                \vspace{-10pt}
            \end{compactitem} \\
            
        11:45am-noon PST & \emph{15min Break} \\
        
        \rowcolor[HTML]{EEEEEE}
        noon-12:11pm PST & \makecell[l]{\textbf{ExaWorks Introduction} \\
        \emph{\small Shantenu Jha (Brookhaven National Laboratory / Rutgers University)} \\
        \emph{\small Kyle Chard (University of Chicago / Argonne National Laboratory)}} \\
        
        12:10-1:00pm PST & \textbf{Review Roadmap Developed by Workflows User and Developer Communities} \\
        
        \bottomrule
    \end{tabular}
\end{table}